\documentclass{sf2a-conf}
\usepackage{graphicx}

\begin{document}
\TitreGlobal{SF2A 2008}
\title{The Chinese-French SVOM mission for Gamma-Ray Burst studies}
\author{Basa, S.}  \address{Laboratoire d'Astrophysique de Marseille, Marseille, France. \\Email: stephane.basa@oamp.fr}
\author{Wei, J.}  \address{National Astronomical Observatories of China, Beijing, China.}
\author{Paul, J.}  \address{AstroParticule et Cosmologie, Paris, France.}
\author{Zhang, S. N.}  \address{Institute of High Energy Physics, Beijing, China.}
\author{the SVOM collaboration}  \address{France: APC Paris, CEA-IRFU Saclay, CESR Toulouse, CNES, IAP Paris, INSU, LAM Marseille, LATT Toulouse, OHP Saint-Michel; China: CAS, CNSA, IHEP Beijing, NAOC Beijing, SECM Shanghai, XIOPM Xi'an; Italy: INAF-IASF Milano.}
\runningtitle{The Chinese-French SVOM mission}
\setcounter{page}{1} 

\maketitle
\begin{abstract}
We present the Space-based multi-band astronomical Variable Objects Monitor mission (SVOM) decided by the Chinese National Space Agency (CNSA) and the French Space Agency (CNES). The mission which is designed to detect about 80 Gamma-Ray Bursts (GRBs) of all known types per year, will carry a very innovative scientific payload combining a gamma-ray coded mask imagers sensitive in the range 4 keV to 250 keV, a soft X-ray telescope operating between 0.5 to 2 keV, a gamma-ray spectro-photometer sensitive in the range 50 keV to 5 MeV, and an optical telescope able to measure the GRB afterglow emission down to a magnitude limit M$_R=23$ with a 300 s exposure. A particular attention will be also paid to the follow-up in making easy the observation of the SVOM detected GRB by the largest ground based telescopes.

Scheduled for a launch in 2013, it will provide fast and reliable GRB positions, will measure the broadband spectral energy distribution and temporal properties of the prompt emission, and will quickly identify the optical afterglows of detected GRBs, including those at very high redshift.
\end{abstract}

\section{Introduction}

The study of GRBs has the potential to expand and revolutionize our understanding of key astrophysical issues. In the coming years they will undoubtedly shed new light on the evolution of the young universe, particularly on the history of star formation, the metal enrichment of galaxies, and the re-ionization of the intergalactic medium. In parallel they will bring crucial insights on the mechanisms driving supernova explosions, the radiation processes at work in regions of space containing a huge energy density, and will provide reliable triggers for gravitational waves and high energy neutrinos detectors. 

The recent interest in GRBs is illustrated by the growing number of instruments on ground and in space dedicated to their studies which are now under construction or under study. In order to fulfil these scientific promises, future studies must rely on the availability of a continuous flow of accurate positions, but also on the measure of many additional parameters (e.g. redshift, E-peak, jet break time,  ...), which are crucial for the understanding of the phenomena itself and for their use as cosmological probes. 

The Sino-French SVOM mission (Space-based multi-band astronomical Variable Objects Monitor) is conceived to:
\begin{itemize}
\item Permit the detection of all known types of GRBs, with a special care on highly redshifted (z$>$6) and sub-luminous GRBs.
\item Measure the temporal properties and the broadband spectral shape of the prompt emission, from visible to MeV domains. 
\item Identify quickly the afterglow emission of detected GRBs, from visible to X-rays domains.
\item Measure the temporal evolution and the broadband spectral shape of the early and late afterglow emission, from visible to X-rays domains.
\item Provide fast and reliable positions, and redshift indicators of detected GRBs.
\end{itemize}

\section{The SVOM mission}
\subsection{The concept}

The constant advances in the field of GRB studies is made possible by the increasing synergy between space and ground instruments. The SVOM mission has been designed to optimize particularly this synergy. The on-board instruments will permit the detection of the GRBs, their localization from arcminutes to arcsecondes accuracy, the study of the prompt emission, the early detection and follow-up of visible afterglows, and the primary selection of high-redshift candidates (z$>$6). The ground segment will permit the fast distribution of the alerts, the localization of GRBs with sub-arcsecondes precision and the multi-band photometry of the afterglow and the prompt emissions, from the visible to the near-infrared domains. 

These functions will be achieved by a set of wide and narrow field instruments. Such a combination requires a very specific observation scenario, which is based on the successful experience of the Swift mission (Gehrels et al., 2004):
\begin{enumerate}
\item the detection is done by a very wide field gamma-ray imaging instrument able to derive on-board localization with few arcminutes accuracy;
\item the position is immediately transmitted to the ground segment and to the scientific community through a  VHF stations network;
\item in parallel the satellite slews rapidly and automatically (when safely possible, regarding to the pointing constraints) to position the GRB in the narrow field of view of the onboard instruments, an X-ray and an optical telescopes, which will study the afterglow and provide refined coordinates. The observations start less than 5 minutes after the detection. 
\end{enumerate}

The scientific objectives of the mission put a very special emphasis on two categories of GRBs: very distant events at redshift greater than 6, which constitute exceptional cosmological beacons, and faint/soft nearby events, which allow probing the nature of the progenitors and the physics at work in the explosion. These goals have a major impact on the design of the mission: the onboard hard X-ray imager must be sensitive down to 4 keV and able to compute image and rate triggers on-board, and the follow-up telescopes on the ground must be sensitive in the near infrared.

\subsection{The scientific payload}

To achieve all these functions, the SVOM mission will operate a set of four instruments in space that constitute the scientific payload of the satellite.

{\em ECLAIRs} is a 2D-coded mask imager sensitive from 4 to 250 keV, with a field of view (FoV) of $89^\circ \times 89^\circ$ and a localization accuracy better than 10', at 7 $\sigma$ (Mandrou et al., 2008). It will deliver triggers by seeking continuously for the appearance of new transient sources in the hard X-ray energy domain, and by determining their localization on the sky (Schanne et al., 2007). When a new source candidate is detected, an alert will be automatically transmitted to the satellite and to the ground within 1 minute via a VHF network.

{\em GRM} (Gamma-Ray Monitor) is a set of two gamma-ray spectro-photometers sensitive in the range 50 keV to 5 MeV. It will cover the same FoV than ECLAIRs and provide a measurement of the peak energy, E-peak.

{\em XIAO} (X-ray Imager for Afterglow Observations) is a mirror focusing X-ray telescope operating from 0.5 to 2 keV, with a FOV of $23' \times 23'$ and a localization accuracy better than 10'',  at 5 $\sigma$ (Mereghetti et al., 2008). It will reach a sensitivity of about 5-10  $\mu$Crab in 10 ks, at 5 $\sigma$. It will provide an intermediate step between the first localizations at several arcminutes given by Eclairs and the precise localizations that can be achieved with an optical telescope. The refined position will be also transmitted via the alert network.

{\em VT} (Visible Telescope) is a 45 cm visible telescope operating from 400 to 950 nm, with a FOV of  $21' \times 21'$ . It will reach a sensitivity of about 23 magnitudes, in the R band, in a 300 s exposure time, at 5 $\sigma$. Subimages, centered on the position provided by XIAO, will be transmitted to the ground in order to detect quickly an optical emission from the GRB and refine its localization accuracy at less than 1''. This position will be also transmitted via the alert network.

\subsection{The ground segment}

Another key elements of the SVOM mission are the Ground Wide Angle Cameras (GWACs) and the Ground Follow-up Telescopes (GFTs). The GWACs, an array of wide FoV optical cameras operating in the optical domain, will permit a systematic study of the visible emission during and before the prompt high-energy emission. It will cover a field of view of about 8000 deg$^2$, with a sensitivity of about 15 magnitudes at 5 $\sigma$ (under the full Moon condition), in the V band and with a 15 s exposure time. It will monitor continuously the field covered by ECLAIRs  in order to observe the visible emissions of more  than 20 \% of the events, at least 5 minutes before and 15 minutes after the GRB trigger.

The GFTs, two robotic 1-meter class telescopes (one managed by France, an other one by China), will point automatically their field-of view towards the space-given error box within tens of seconds after the alert reception and will provide panchromatic follow-up (visible to near-infrared). They will contribute to the improvement of the link between the scientific payload and the largest telescopes by measuring the celestial coordinates with an accuracy better than 0.5'', and by providing an estimate of its photometric redshift in less than 5 min after the beginning of the observations. These information will be available to the scientific community through an alert message. Evenly placed on the Earth (one in South America in a place to be defined, the other one in China), they will be in a position to start the research of the GRB optical emission immediately after the alert reception in more than 40 \% of the cases. 

All the alerts of any new transient candidate will be transmitted from the scientific payload to the ground  in real-time via a VHF real-time network, which is based on the successful experience of the Hete-II mission\footnote{The main difference is due to the higher orbit inclination foreseen for SVOM satellite, which implies to spread VHF stations between $\pm30^\circ$ of latitudes.} (Tamagawa et al., 2003). The prompt alert will be distributed to the scientific community in the first minutes after the on-board detection through the GCN network. Main characteristics of the burst, the useful ones for follow-up campaigns, will be determined from a subset of data downlinked in real-time through the VHF network, before the full data are available through the X-band.

\subsection{The follow-up program}

Follow-up telescopes will play a decisive role in the scientific return of the SVOM mission by extending its capabilities to domains not covered by the scientific payload and the ground segment: deep near-infrared photometry, spectroscopy, polarimetry, large wavelength coverage (radio to Ultra High Energy), ... 

A follow-up program will be systematically organized and  will have to guarantee a uniform quality to the largest possible sample. Possible instruments which could be involved on the SVOM follow-up are:

{\em Radio:}   ALMA, VLA, ...

{\em Visible  and infrared:}  Tarot, Raptor, 2.2m MPE, Falkes Telescopes, Liverpool Telescope, ... (robotic telescopes) and VLT (Hawk-I, XShooter), JWST, Subaru, ... (large telescopes).

{\em Gamma-ray:}  Swift (if still in operation), Fermi, ...

{\em UHE:}  Antares, Auger, CTA, Hess, Ice Cube, Magic, Milagro,  ...

\subsection{Observing strategy}

The selected SVOM orbit is circular with an altitude of $\sim$600 km and an inclination angle of $\sim$30$^\circ$, with a precession period of 60 days. To allow fast and systematic follow-up observations by ground-based telescopes, the satellite orientation is quasi anti-solar, granting that the bursts are discovered in the night part of the sky. Strong Galactic sources and the Galactic plane where heavy extinction prevents the detection of afterglows are avoided. Thanks to this strategy, about 75 \% of the GRBs will be visible at the time of their detection from at least one of the three major ground-based observatories (Cerro Paranal, Mauna Kea and Roques de los Muchachos). 

A majority of bursts will be immediately observable at the end of the satellite automatic slew maneuver: about 60\% are observable ÒimmediatelyÓ (starting observation 5 minutes after the detection, for at least 5 minutes), the rest being observable between 40 and 60 minutes later due to Earth obstruction, which may occur just few minutes after the trigger (Cordier et al., 2008).

\section{The scientific performances}

The SVOM mission offers a very attractive combination of instruments. The burst observation rate for the Eclairs instrument is estimated to about 80 per year for a 7 $\sigma$ level detection, with about 10 \% of the events at a redshift larger than 6. Despite its smaller effective surface than the Burst Alert Telescope of Swift,  Eclairs holds greater potential for the discovery of highly redshifted and faint gamma-ray bursts thanks to a low-energy threshold of 4 keV.  Simultaneously, the GRM will be able to provide systematically a precise estimation of the peak energy parameter and the GWACs an observation of the prompt emission in the visible domain. 

After an automatic satellite slew maneuver lasting less than 5 minutes, XIAO, with a sensitivity close to the Swift X-Ray Telescope, and the VT, with  a sensitivity significantly improved over the Swift UVOT, completed by the two GFTs on ground, will insure a systematic multi-wavelength follow-up for several hours. In particular the VT will allow the detection of nearly 75\% of GRBs in the visible domain, during the first orbit (Akerlof \& Swan) and, for the first time, to explore the realm of ``dark GRBs''. All these instruments are included in a chain allowing a refinement of the localization, from few arcminutes  to sub-arcsecondes. All these alerts will be distributed to the scientific community in real time through the GCN alert network.

A significant fraction of the time will be also available for non-GRB science like the discovery or the follow-up of cataclysmic variables, active stars, active galactic nuclei, supernovae, ... It will be possible to propose a program consisting in an observation of a given target for typically few consecutive orbits (programs maximizing the simultaneous use of several instruments will be always favored). Such a program will have always a lower priority than the GRB core program:  the capabilities to detect a GRB must not be affected and the observation is stopped whenever a GRB is detected and will only resume after completion of the follow-up campaign. Target of Opportunity (ToO) observations will be also accepted for unpredictable events discovered in the routine scrutiny of the SVOM data or proposed by external partners.

\section{Conclusion}
SVOM is a very ambitious multi-wavelength mission operating from the optical to the gamma-ray domains. It is designed to detect all known types of GRBs, to provide fast and reliable positions, to measure the broadband spectral shape and temporal properties of the prompt emission, and to quickly identify the visible and X-ray afterglows of detected events, including those ay very high redshift.

Scheduled for a launch in 2013 and for at least 3 years, it will open the door to systematic and accurate GRB studies, allowing a better understanding of the phenomenon and of the Universe!

\section*{Acknowledgement}
The french contribution to the SVOM mission is completed under the responsibility and leadership of the Centre National d'{}Etudes Spatiales (CNES). The french authors are grateful to CNES support.


\begin{thebibliography}{}
\bibitem{Akerlof} Akerlof  \& Swan 2007, ApJ, 671, 1868.
\bibitem{Cordier} Cordier, B., 2008, Proc. Gamma-Ray Bursts 2007, Santa Fe, USA, AIP Conf. Proc., 1000, 585-588.
\bibitem{Gehrels} Gehrels, N. et al, 2004, ApJ 611, 1005-1020.
\bibitem{Mandrou}  Mandrou, P. et al., 2008, Proc. Nanjing GRB Conference, Nanjing, China, 2008arXiv0809.1057M.
\bibitem{Mereghetti} Mereghetti, S. et al., 2008, Proc.SPIE Astronomical Telescopes and Instrumentation 2008, Marseille, France, 2008arXiv0807.0893M.
\bibitem{Schanne} Schanne, S. et al., 2007, Proc. 30$^{th}$ International Cosmic Ray Conference, Merida, Mexico, 2007arXiv0711.3754S.
\bibitem{Tamagawa} Tamagawa, T. et al., 2003, Proc. 28$^{th}$ International Cosmic Ray Conference, Trukuba, Japan.

\end{thebibliography}
\end{document}